\documentclass[aps,prb,preprint,superscriptaddress,floatfix,showpacs]{revtex4}
\usepackage{graphicx}

\begin{document}


\title{Quantum critical metamagnetism of  Sr$_3$Ru$_2$O$_7$ under hydrostatic pressure}


\author{W. Wu}
  \affiliation{Department of Physics, University of Toronto, 
     60 St.\ George Street, Toronto, Canada M5S 1A7}
\author{A. McCollam}
  \affiliation{Radbout University Nijmegen, High Field Laboratory, 
     Faculty of Science, 
     P.O.\ Box 9010, 6500 GL Nijmegen, The Netherlands}
\author{S. A. Grigera}
  \affiliation{Scottish Universities Physics Alliance, School of 
       Physics and Astronomy, University of St.\ Andrews, North Haugh, 
       St. Andrews  KY16 9SS, UK}
\author{R. S. Perry}
  \affiliation{Center for Science at Extreme Conditions, School of Physics, 
        The University of Edinburgh, Edinburgh, EH9 3JZ, Scotland}
\author{A. P. Mackenzie}
  \affiliation{Scottish Universities Physics Alliance, School of 
       Physics and Astronomy, University of St.\ Andrews, North Haugh, 
       St. Andrews  KY16 9SS, UK}
 \affiliation{Canadian Institute for Advanced Research, 
    Quantum Materials Program, 
    180 Dundas St.\ W., Suite 1400, Toronto, ON, Canada M5G 1Z8}
\author{S. R. Julian}
  \email[]{sjulian@physics.utoronto.ca}
  \affiliation{Department of Physics, University of Toronto, 
     60 St.\ George Street, Toronto, Canada M5S 1A7}
 \affiliation{Canadian Institute for Advanced Research, 
    Quantum Materials Program, 
    180 Dundas St.\ W., Suite 1400, Toronto, ON, Canada M5G 1Z8}

\date{\today}

\begin{abstract}
Using ac susceptibility, we have determined the pressure dependence of 
  the metamagnetic critical end point temperature $T^*$ for field applied 
  in the $ab$-plane in the itinerant metamagnet $\rm Sr_3Ru_2O_7$. 
We find that $T^*$ falls monotonically to zero as pressure increases, 
  producing a quantum critical end point (QCEP) at 
  $P_c\sim 13.6\pm 0.2$~kbar. New features are observed near the 
  QCEP -- the slope of $T^*$ vs pressure changes at $\sim$12.8~kbar, 
  and weak subsidiary maxima appear on either side of the main susceptibility 
  peak at pressures near $P_c$  -- indicating that some new physics comes 
  into play near the QCEP. 
Clear signatures of a nematic phase, however, 
  that were seen in field-angle tuning of $T^*$, are not observed.  
As $T^*$ is suppressed by pressure, the metamagnetic peak in the 
  susceptibility remains sharp as a function of applied magnetic field. 
As a function of temperature, however, the peak becomes broad with only a 
  very weak maximum, 
  suggesting that, near the QCEP, the uniform 
  magnetization density is not the order parameter for the metamagnetic 
  transition. 

\end{abstract}



\maketitle


\section{Introduction}

Quantum criticality continues to attract a lot of interest, much of it in 
  connection with its role in generating exotic behavior of correlated 
  electron systems. 
The  original model of a quantum critical point involved a second-order 
  phase transition being shifted to 0~K by some non-thermal 
  tuning parameter such as pressure, chemical doping or magnetic field 
  \cite{Hertz76}. 
The $T\rightarrow 0$ critical point, i.e.\ the quantum critical point (QCP), 
  gives rise to nontrivial emergent excitations 
  that control the physics over a significant portion of the phase diagram. 
In metals, electrons show non-Fermi liquid behavior in the quantum critical 
  region, but also, near the QCP, electrons show a strong tendency to 
  re-organize themselves into new stable phases such as exotic 
  superconducting states.

Recently, a new kind of quantum critical point, 
  associated with a first-order metamagnetic phase transition (MMT) 
  in which no symmetry is broken, 
  has been observed in $\rm Sr_3Ru_2O_7$. 
Metamagnetism is empirically defined as a superlinear change of 
  magnetization vs magnetic field in a narrow field range 
  (a discontinuous jump in magnetization in the case of a first-order MMT).  
Quantum criticality is achieved by suppressing the end point of this 
  first-order phase transition to absolute zero \cite{Grigera01}. 
The term ``quantum critical end point'' (QCEP) is used to distinguish 
  this from a QCP that involves symmetry breaking. 

 \begin{figure}
\centering
 \includegraphics[width=0.45\textwidth]{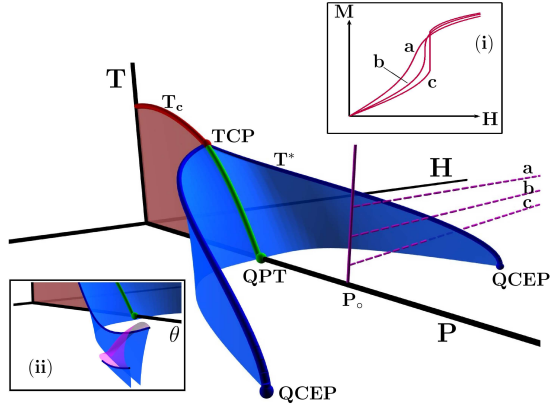}
 \caption{ Main figure: the proposed generic phase diagram of a metal near the 
             border of ferromagnetism \cite{Pfleiderer01,Griffiths67,Belitz02}.
             As the ferromagnetic transition temperature $T_c$ is suppressed 
             by a control parameter $P$, it changes from second- to 
             first-order at a tricritical point, TCP. From the line of 
             first-order transitions connecting TCP with the first-order 
             quantum phase transition, QPT, two metamagnetic `wings' emerge 
             (blue surfaces), corresponding to surfaces in ($T$,$P$,$H$) space 
             at which the magnetization jumps discontinuously (see inset (i)). 
             The line of critical end points, $T^*$, goes to 0~K at the 
             quantum critical end-point, QCEP. In ultra-pure $\rm Sr_3Ru_2O_7$,
             as $T^*$ is tuned by the angle of the magnetic field, the QCEP 
             does not appear.   Instead, a nematic phase is found, enclosed on 
             the sides by two first-order metamagnetic jumps, and on top by a 
             probable second-order phase boundary (inset (ii)). 
\label{genericpha}}
\end{figure}

Figure \ref{genericpha} shows the suggested `generic' phase diagram of a 
  metal on the border of ferromagnetism 
  \cite{Pfleiderer01,Griffiths67,Belitz02}. 
It has been applied, for example, to $\rm CoS_2$ \cite{Goto01}, MnSi 
  \cite{Pfleiderer01}, $\rm CeRu_2Si_2$ \cite{Flouquet95} and 
  UGe$_2$ \cite{Taufour10}. 
In this model, a second order phase transition to 
  a spontaneously ordered ferromagnetic state occurs at $T_c$ at $H=0$. 
$T_c$ is then suppressed by a tuning parameter such as hydrostatic 
  pressure, but as $T_c$ falls, it encounters a tricritical point, TCP, 
  at which the second-order transition becomes first-order. 
At the tricritical point, two metamagnetic `wings' emerge 
  (at positive and negative magnetic field), 
  representing surfaces at which  there is a first-order metamagnetic 
  jump in the magnetization as a function of applied magnetic field $H$. 
The top of the wings is delimited by a line of critical points $T^*(H,P)$, 
  which separates the first-order jump from a continuous super-linear 
  crossover behavior in the $M$ vs $H$ curve. 
This is illustrated in Figure \ref{genericpha}(i): as $H$ is increased 
  along an isotherm with $T < T^*$, represented by the dashed line 
  labelled  $c$, the magnetization jumps discontinuously when the line 
  passes through the surface; 
  alternatively, if $T>T^*$, as in line $a$, there is no discontinuity,  
  only a crossover. 
At T* the magnetic susceptibility, $\chi = dM/dH$, should diverge. 
The point on the phase diagram at which $T^*\rightarrow 0$~K is the 
  quantum critical end-point \cite{Grigera01}.

There is considerable interest in the behavior near the quantum critical 
  end point in $\rm Sr_3Ru_2O_7$ \cite{Grigera04, Borzi07,Rost09}. 
At ambient pressure, for magnetic fields applied parallel to the 
  $ab$-plane so that the magnetic-field angle, $\theta$, is equal to zero, 
  $\rm Sr_3Ru_2O_7$ is believed  to lie on the generic phase diagram 
  roughly where the dashed lines, labelled $a$, $b$ or $c$, are situated 
  in Figure \ref{genericpha}. 
That is, the ground state of Sr$_3$Ru$_2$O$_7$ is paramagnetic, but it is very 
  close to being ferromagnetic, as demonstrated by the fact that, 
  while highly hydrostatic pressure drives 
  Sr$_3$Ru$_2$O$_7$ {\em away} from ferromagnetism \cite{Ikeda01,Sushko04} 
  and causes the metamagnetic transition field to increase \cite{Chiao02}, 
  uniaxial stress applied in the $c$-axis direction \cite{Ikeda01,Ikeda04} 
  drives the system {\em to} ferromagnetism at very low uniaxial stresses of 
  around 1 kbar. 
 (Note that the first high-pressure study of Sr$_3$Ru$_2$O$_7$ inadvertently 
  had a large uniaxial stress component and produced ferromagnetism  
  around 10 kbar \cite{Ikeda00}.) 
In an applied magnetic field, 
  rotating the field away from the $ab$-plane to the 
  magnetically harder $c$-axis 
  seems to be equivalent to tuning away from ferromagnetic order: 
  $T^*$ falls, and a study of $T^*$ vs $\theta$ for ``high-purity'' 
  single crystals (having residual resistivity $\rho_o\sim 2.4~\mu\Omega$~cm) 
  shows that the QCEP, $T^*\rightarrow 0$~K, occurs at about 
  $\theta=80^\circ$ \cite{Grigera03}.

In even higher purity samples, however, having $\rho_o<0.5~\mu\Omega$~cm and referred to in this paper as ``ultra-pure'', $T^*$ does not go to zero as  a function of $\theta$, rather
it has a minimum around $\theta \sim 60^\circ$, and then rises again accompanied by
another, nearby, first-order jump at slightly higher field. This is illustrated
 schematically  in
Figure \ref{genericpha}(ii). It has been shown that these two first-order transitions enclose a novel ‘nematic’  phase (the region under the pink dome in Figure \ref{genericpha}(ii))  with strongly anisotropic transport properties that break the symmetry of the lattice \cite{Grigera04,Borzi07}. 
The
nature of the nematic phase is not well understood, but it has been speculated that
the nematic phase maybe a result of a d-wave distortion of the Fermi surface
arising from a Pomeranchuk instability \cite{Grigera04, Kee05, Raghu09}. 
Recently, it was proposed
that the nematic phase is a spatially modulated magnetic state analogous to a
Fulde-Ferrell-Larkin-Ovchinnikov (LOFF) phase \cite{Berridge09, Berridge10}.

Prior to $\rm Sr_3Ru_2O_7$, metamagnetic transitions had been reported in several other
d- or f-electron metals such as UPt$_3$ \cite{Frings83} 
and URu$_2$Si$_2$ \cite{Visser87}. 
However, only in
$\rm Sr_3Ru_2O_7$ has it been possible to study the quantum critical end point, and these
studies have been limited to field-angle tuning as described above. Field-angle
tuning has been proposed to play a role analogous to pressure, based on the
assumption that the field-angle suppresses the metamagnetism through angle-dependent
magnetostriction \cite{Grigera03}. 
In this sense, the phase diagram with the field-angle
as tuning parameter could have a close relation to the pressure-induced
phase diagram obtained from Ginzburg-Landau treatments \cite{Pfleiderer01, Grigera02}. 
However,
in changing the field-angle the symmetry also changes, and nematic signatures
are strongest when the symmetry is high, i.e.\ when the field is close to either
the $c$-axis or the $ab$-plane \cite{Borzi07}. 
A different explanation of the role of the field-angle,
suggested by Raghu et al.\ \cite{Raghu09} 
and Berridge et al.\ \cite{Berridge10}, 
is that field-angle
moves the system through the phase diagram via orbital effects, i.e.\ by
modification of the band structure through the spin-orbit and orbital-Zeeman
coupling \cite{Raghu09}. 

This change of symmetry and orbital coupling as the direction of the 
  field is changed in field-angle tuning complicates the interpretation of 
  the results. 
If the metamagnetic transition were tuned with pressure then the 
  symmetry and angle-dependent orbital coupling would not change, and this  
provides strong motivation for exploring the
metamagnetic quantum criticality of Sr$_3$Ru$_2$O$_7$ under hydrostatic pressure. An
intriguing question is whether the new nematic phase appears with pressure
tuning.

In this paper we report an investigation, using ac-susceptibility under hydrostatic pressure,  
  of the metamagnetic quantum criticality of 
  ultra-pure crystals of Sr$_3$Ru$_2$O$_7$ 
  for fields applied in the $ab$-plane. 
Compared to $H\parallel c$ where the nematic phase has already been observed, 
  using $H\parallel(ab)$ has the disadvantage that 
  the magnetic field breaks the in-plane symmetry; 
  however we wished, in this first study at least, 
  to follow the evolution of the critical end-point as a function of pressure, 
  and this is not possible for $H \parallel c$ because 
  the field-angle has already tuned the system to the quantum 
  critical region even at zero pressure. 
We note that weak signatures of nematicity have been reported for 
  $H \parallel (ab)$, although not at the primary metamagnetic transition \cite{Perry05}. 
We found that $T^*$ decreases monotonically with increasing pressure, going rather suddenly to zero above 12.8~kbar. 
The QCEP occurs at $P_c\sim 13.6\pm 0.2$~kbar. 
We also  observed  that  the divergence of the susceptibility at $T^*$, 
  illustrated by the slope of curve (b) in Figure \ref{genericpha}(i), 
  weakens dramatically as $P_c$ is approached, suggesting that the naive picture of 
  metamagnetism as field-induced ferromagnetism may not apply to Sr$_3$Ru$_2$O$_7$ 
  near the QCEP; rather it may arise from the suppression of 
  antiferromagnetic correlations, or a change in some higher-order correlation 
  function of the electron system.

\section{Experiment}

Hydrostatic pressure was applied using a BeCu clamp cell. To achieve a highly
homogeneous pressure, Daphne oil 7373 was used as the transmitting medium. 
The pressure at low temperatures was determined from the known pressure 
    dependence of the superconducting transition temperature of tin.
The
ac susceptibility was measured using a set of detection coils and a drive coil.
The detection coil set is comprised of three coils, with the central coil connected
antiparallel to the two end coils. The drive coil is concentrically wound around
the three pick-up coils. This configuration significantly reduces background
pick-up from the feedthrough that carries the wires into the high pressure region,
allowing us to see the metamagnetic peak more clearly. A low frequency
excitation field of 14 Hz,  generated by the ac current in the drive coil,  was employed
in order to reduce finite-frequency effects \cite{Duyneveldt82}. At 13.4 kbar, 83 Hz was
also used in order to test for frequency dependence. A sample with approximate  dimensions 0.7$\times$0.7$\times$1.7~mm$^3$ was placed in the central pick-up coil and
thermally grounded to the mixing chamber  through silver
and copper wires. The response of the sample was detected by a lock-in
amplifier, preceded by a low temperature transformer with a turns ratio of $\sim$100
and a $\times$1000 low-noise pre-amplifier. The sample used here was cut from  
an ultra-pure single 
crystal of Sr$_3$Ru$_2$O$_7$ grown at St.\ Andrews University, UK. The
residual resistivity was measured to be $\rho_{res} < 0.5 ~\mu\Omega$~cm.

For all the ac susceptibility measurements, 
the samples were cooled in zero field and 
the dc field was applied in the $ab$-plane, 
i.e.\ parallel to the ac field. 
The sweep rate of the dc field was 0.02 T/min, the fastest rate for which there was 
  no sign of heating in the lowest-temperature data.  
At pressures below 12.8 kbar 
  we used only data from downsweeps,  
  whereas at 12.8 kbar and above we averaged the results of up and down sweeps. 
At the sweep rate of 0.02 T/min
  we did not resolve any hysteresis in the positions of the peaks 
  between up and down sweeps, beyond the lag that is 
  expected from the time-constants of our measurement system. 
(Unambiguous evidence for hysteresis is, however, supplied by the
    presence of a peak in the imaginary part of the susceptibility, 
    which will be described below.) 
In averaging up and down sweeps, as was done at 12.8 kbar and above, we first
    shifted the field-axes by the tiny amount required to make positions of the peaks match.

In this investigation, we are only interested
in the relative variation of the ac susceptibility due to the metamagnetic transition
($\Delta\chi$), so a slowly varying background signal
including the paramagnetic susceptibility of 
Sr$_3$Ru$_2$O$_7$ has been subtracted using
a 5th degree polynomial fit. The amplitude of the ac modulation field was
approximately 0.1 G. The absolute ac susceptibility was left unresolved and
therefore arbitrary units, a.u., are used in all the figures, however the 
relative amplitude of the peaks at different pressures can be compared 
directly, as the same modulation amplitude and frequency, and the same 
electronics, were used throughout.

\begin{figure}
\includegraphics[width=0.45\textwidth]{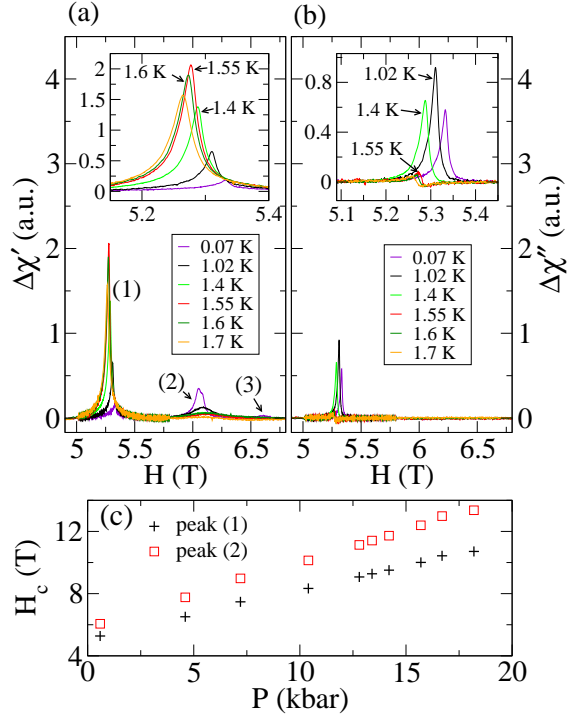}
\caption{The real (a) and imaginary (b) parts of the ac magnetic susceptibility
of Sr$_3$Ru$_2$O$_7$ at 0.59~kbar as the in-plane 
dc field is swept through the metamagnetic
transitions. The data are labelled as $\Delta \chi^\prime$ and $\Delta \chi^{\prime\prime}$ respectively; a slowly
varying background has been subtracted. Although we use arbitrary units, the
same coil and sample are used in all measurements so relative amplitudes at different pressures can
be compared.  
Three successive peaks are observed in the susceptibility, numbered 
 (1), (2) and (3) in Figure (a). 
The inset in each panel shows an 
 expanded plot around peak (1), 
 which is the focus of this paper.  For peak (1), with decreasing 
 temperature from 1.7~K, $\Delta \chi^\prime_{max}$  initially grows, reaches 
 a maximum at 
 $T^* = 1.55$~K, 
 and then decreases as the temperature is further 
 reduced. 
(b) the peak in $\Delta\chi^{\prime\prime}$ only starts to appear 
 below 
 $T^* = 1.55$~K, 
 and then increases rapidly in amplitude as the temperature 
 is reduced. 
No signal in $\Delta\chi^{\prime\prime}$  is observed at the positions 
 of peaks (2) or (3). The small step in $\Delta \chi^{\prime\prime}$ at 
 temperatures above 1.55~K may be the result of changing eddy currents 
 in the sample.
%
Figure (c) shows the pressure dependence of the critical metamagnetic field $H_c$ at 
   $T^*$, as a function of pressure for peaks (1) and (2).  For pressures above 
   13.4~kbar, $H_c$ at $\sim 0.07$~K is used.  
%
\label{0.59kbar_realandimag}}
\end{figure}

\section{ Results}

Figure \ref{0.59kbar_realandimag} shows the ac susceptibility of Sr$_3$Ru$_2$O$_7$,  $\Delta\chi$,  
  as a function of 
decreasing
 dc field
  under a hydrostatic pressure of 0.59~kbar. 
The real part of the ac susceptibility, $\Delta\chi^\prime$, exhibits a pronounced peak, 
  numbered (1) in Figure \ref{0.59kbar_realandimag}, across the metamagnetic
  transition at a field $H_M \approx 5.3$~T, and two minor peaks, numbered (2)
  and (3), at higher fields, $H \approx 6.06$~T and $H \approx 6.6$~T. 
These features are believed to reflect sharp peaks in the density of states, 
  such as would arise for example from a van Hove singularity \cite{Grigera04,Kee05}, 
  but a detailed connection with the rather complex electronic structure 
  of Sr$_3$Ru$_2$O$_7$ \cite{Mercure10} has not yet been possible. 
Peak (2) at $H \approx  6.06$~T
  evolves into a double feature with decreasing temperature, reminiscent of the
  static differential susceptibility reported for this peak in Ref. \cite{Perry05}. 
As can be seen from Figure \ref{0.59kbar_realandimag}(c), using data
  described below we followed peaks (1) and (2) up to 18 kbar,
  finding that both peaks shift to higher field roughly linearly 
  with increasing pressure. Peak (1) increases with pressure at a rate of 
  0.3~T/kbar up to 18.2~kbar, while $H_c$ for peak (2) rises somewhat 
  faster: the separation between peak (1) and peak (2) expands from 
  0.79~T at 0.59~kbar to 2.63~T at 18.2~kbar.
The size of peak (2) depends 
  more weakly on pressure and temperature than that of peaks (1) and (3), 
  and in fact peak (3) disappears quickly with 
  rising temperature and pressure. 
Within the temperature and pressure range studied 
  we were unable to resolve any imaginary part of the susceptibility for 
  either peak (2) or peak (3). 

For peak (1), Figure \ref{0.59kbar_realandimag}(a) shows that the peak in 
$\Delta \chi^\prime$ reaches its maximum
at 1.55~K, while Figure \ref{0.59kbar_realandimag}(b)  
shows that the corresponding imaginary part $\Delta \chi^{\prime\prime}$ of the
ac susceptibility starts growing only below 1.55~K. This behavior arises from a 
first-order MMT transition terminating in a
critical point at a temperature $T^*\sim 1.55$~K \cite{Grigera03}: 
  above $T^*$,  the $M$ vs $H$  curve is a crossover 
  that sharpens as $T \rightarrow T^*$; below $T^*$, the dynamical
  response becomes sensitive to the physics of a first-order metamagnetic phase
  transition, such as domain wall movement, so that  the real part of the ac
  susceptibility decreases while the imaginary part grows. 
It is also observed that
  the metamagnetic critical field has a weak temperature dependence, decreasing
  by 0.074~T from 0.1~K to 1.8~K.

\begin{figure}
\includegraphics[width=0.45\textwidth]{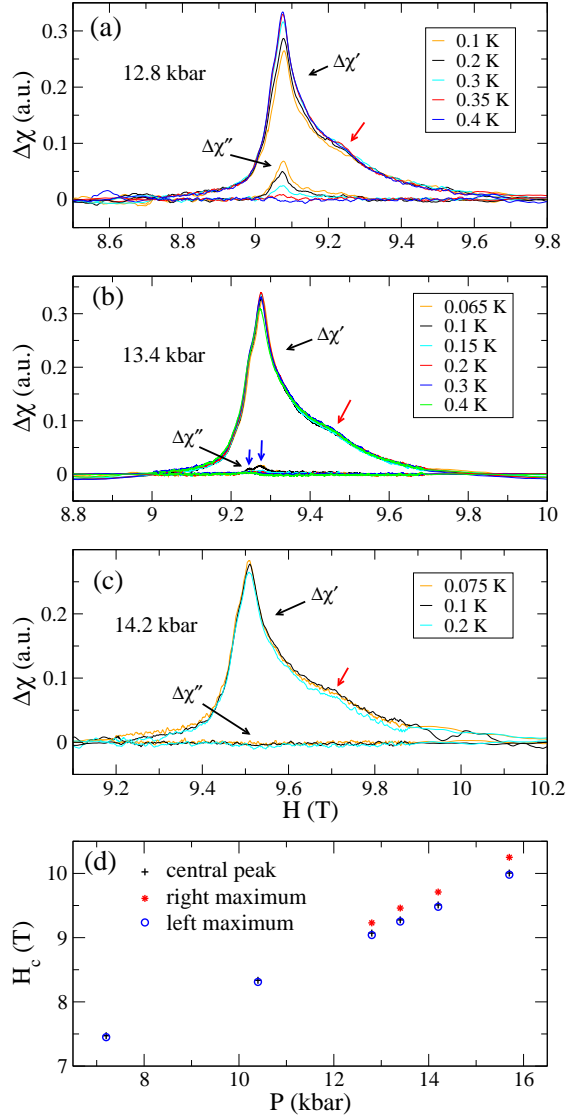}
\caption{The temperature evolution of the ac susceptibility 
  across the metamagnetic transition for 12.8~kbar (a), 13.4~kbar 
  (b) and 14.2~kbar (c). 
A secondary maximum to the right of the central peak of $\Delta \chi^\prime$, is
marked by the red arrow. $\Delta \chi^{\prime\prime}$ shows a clear peak below 0.3~K at 12.8~kbar, a
much weaker peak below 0.15~K at 13.4~kbar, and no peak down to 0.07~K
at 14.2~kbar. At 13.4~kbar, a double peak feature can be seen in $\Delta\chi^{\prime\prime}$. Note
that the scales on both the vertical and horizontal axes are different for the
three graphs (a), (b) and (c).  (d) shows the critical metamagnetic field $H_c$ at $~T^*$ as a 
function of pressure for peak (1) and 
of the two secondary maxima in $\Delta \chi^\prime$ at $\sim$0.07~K. 
For pressures above 13.4 kbar, $H_c$ at $\sim$0.07~K is used, as in Figure 
\ref{0.59kbar_realandimag}(c).
\label{12.8_13.4_14.2kbar}}
\end{figure}

Data such as that shown in Figure \ref{0.59kbar_realandimag}  
  has been collected at 0.59~kbar, 4.6~kbar, 7.2~kbar, 10.4~kbar, 
  12.8~kbar, 13.4~kbar, 14.2~kbar, 15.7~kbar, 16.7~kbar and 18.2 kbar. 
As pressure increases from 0.59~kbar,
  the critical temperature $T^*$
  decreases, while $H_M$ moves toward higher field. 
As shown in Figure \ref{12.8_13.4_14.2kbar}(a), by
  12.8~kbar, $T^*$ has fallen to 0.375$\pm$0.025~K. 
At this pressure new structure has appeared both above and below the main 
  peak in $\Delta \chi^\prime$.  To the right there is a pronounced bump, 
  or secondary maximum, in $\Delta\chi^\prime$, indicated by the red arrow 
  in Figure \ref{12.8_13.4_14.2kbar}(a).  
$\Delta \chi^{\prime\prime}$ extends asymmetrically out to this 
  secondary maximum.  Similarly, just below the main peak a weak 
  secondary maximum is seen in both $\Delta \chi^\prime$ and 
  $\Delta \chi^{\prime\prime}$.

At 13.4 kbar, $T^*\sim 0.15$~K, and the secondary maxima become more clear in 
  comparison with 12.8~kbar. 
The dissipation signal
corresponding to the central peak in $\Delta\chi^\prime$ diminishes but is still visible; by 13.4~kbar, it has evolved into two distinct peaks (see the  blue arrows in Figure \ref{12.8_13.4_14.2kbar}(b)).     
The left peak in $\Delta\chi^{\prime\prime}$ matches the 
     secondary maximum just below the main peak in $\Delta\chi^\prime$, 
     however $\Delta\chi^{\prime\prime}$ is zero, within our resolution, 
     at the secondary maximum on the right.

At 14.2 kbar (see Figure \ref{12.8_13.4_14.2kbar}(c)), 
  $T^*$ has fallen below 0.07~K, the lowest temperature  reached in
  these measurements.
$\Delta\chi^{\prime\prime}$ remains flat down to 0.07~K, showing that the peaks in
$\Delta\chi^\prime$ are crossovers. The secondary maxima to the right and left of the central maximum  in $\Delta\chi^\prime$ 
are still discernible at this pressure.

Figure \ref{12.8_13.4_14.2kbar}(d) zooms in on the portion of 
  Figure \ref{0.59kbar_realandimag}(c) close to $P_c$, showing the 
  shift with pressure of the central peak and the two 
  secondary maxima. 
It can be seen that the features all shift together, and there is no 
  visible change in slope at $P_c$.

\begin{figure}
 \includegraphics[width=0.45\textwidth]{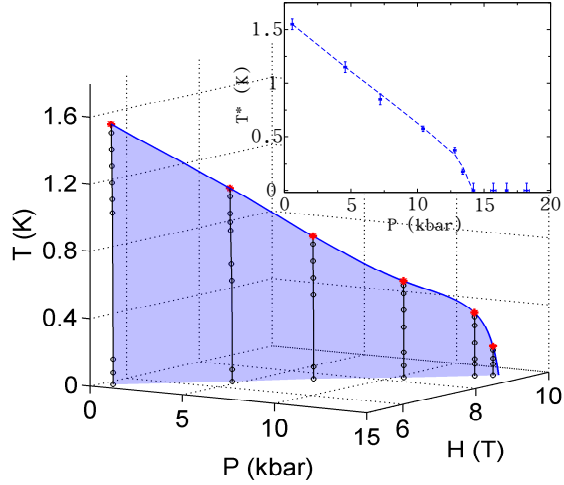}
 \caption{The phase diagram inferred from susceptibility measurements. The
blue and black solid lines are splines of the measured critical end points 
$T^*$ (red) and the position of the MMT below $T^*$ as a function of temperature 
    and field (black) at each pressure, respectively. The
inset shows the projection of the line of critical end points in the ($P$,$T$) plane. For
pressures larger than 14.2~kbar, 0.07~K is taken as the error bar for the critical
temperatures because that was the lowest temperature reached. The quantum
critical end-point is close to 13.6~kbar. The dashed line in the inset is a guide
to the eye.
\label{phasedia}}
\end{figure}

The ($T^*$,$P$,$H$) phase diagram is given in Figure \ref{phasedia}. This represents our measurement
of the tip of  a metamagnetic ‘wing’ that is  shown schematically in Figure \ref{genericpha}.  The critical
temperature $T^*$ falls uniformly from $\sim$1.55~K at $\sim$0.59 kbar to $\sim$0.375~K
at $\sim$12.8~kbar; then $T^*$ drops quickly to below 0.07~K, the lowest temperature reached in these measurements. In the inset, the error bars at pressures above 14.2~kbar extend from
zero to $\sim$0.07~K, but it is reasonable to assume that $T^*$ has fallen to zero at
approximately 13.6~kbar, making this the quantum critical end-point pressure,
$P_c \sim 13.6 \pm 0.2$~kbar. Above $P_c$, the peak in $\Delta\chi^{\prime\prime}$  has disappeared, while the central
peak in $\Delta \chi^\prime$ persists. The secondary maximum above the main peak weakens as the pressure 
    is further increased, and disappears  at $\sim$16.7~kbar.

Figures \ref{0.59kbar_realandimag} and \ref{12.8_13.4_14.2kbar} 
  show $\Delta\chi^\prime$ vs $H$ sweeps at constant temperatures.  
Comparing these figures, we observe the surprising result that although the metamagnetic peak has a strong temperature dependence 
near $T^*$ at low pressures (Figure \ref{0.59kbar_realandimag}),  
for pressures  near $P_c$ (Figure \ref{12.8_13.4_14.2kbar}) this has 
become very weak.

\begin{figure}
 \includegraphics[width=0.45\textwidth]{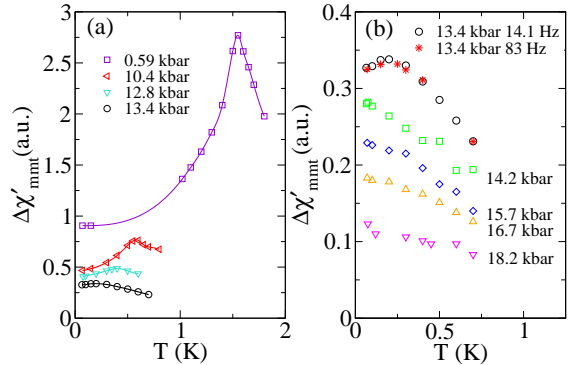}
 \caption{The magnitude of $\Delta\chi^\prime$ at the metamagnetic transition field, $\Delta\chi^\prime_{mmt}$,
as a function of temperature for several pressures, showing the dramatic fall in
 $\Delta\chi^\prime_{mmt}$ as the quantum critical end point is approached. (a): data below $P_c$;
note that the data below 13.4 kbar have been offset to avoid overlap. (b):
expanded plot of the higher-pressure data; the datasets for 13.4~kbar show that
 frequency has little effect on the temperature dependence  of  $\Delta\chi'_{mmt}$.  
 Note that the  gain-settings for the two datasets at 13.4~kbar are different, so the 83~Hz curve has been rescaled by a multiplicative factor to agree with the 14.1~Hz curve at 0.75~K.  
\label{mag_T}}
\end{figure}

This is emphasized in  Figure \ref{mag_T}, which plots the temperature dependence of the 
  maximum in $\Delta\chi^\prime$. 
Clearly, the peak at $T^*$ collapses drastically with increasing pressure: 
  as $P_c$ is approached, the maximum becomes much weaker, and near the quantum critical
  end point it has nearly disappeared. 
This phenomenon has little frequency dependence: 
  Figure \ref{mag_T}(b) includes data for two different frequencies, 14.1~Hz 
  and 83~Hz, at 13.4~kbar, and the two datasets closely overlap.

\section{Discussion}

We have found that, for $H\parallel  ab$,  
application of hydrostatic pressure produces a
quantum critical end point at 13.6$\pm$0.2~kbar in Sr$_3$Ru$_2$O$_7$. This opens new
avenues for studying quantum criticality and metamagnetism in this material.

 As with field-angle tuning from the $ab$ plane to the $c$-axis, hydrostatic pressure
causes a monotonic increase in the metamagnetic transition field $H_M$ and moves
the system away from ferromagnetic order (see Figure \ref{genericpha}). However, the phase
diagram produced by pressure tuning (see Figure \ref{phasedia}) looks very different from that
produced by field-angle tuning for the same ultra-pure quality crystals \cite{Grigera04, Borzi07}.  
In
the latter case, as the system is tuned away from  ferromagnetism, the QCEP
is avoided due to the appearance of the nematic phase bounded by first-order
metamagnetic jumps, so $T^*$ never goes to zero, rather it has a minimum at
 $\theta \sim 60^\circ$ and then rises again as the nematic phase emerges. With pressure, in
contrast, $T^*$ goes to zero, apparently smoothly.

However, despite the similarity of Figure \ref{phasedia} to the tip of the metamagnetic
wing in the generic phase diagram (Figure \ref{genericpha}), the underlying physics seems to
be quite different. According to  the generic model of quantum critical metamagnetism \cite{Millis02},   the susceptibility should be divergent at $T^*$, but Figure \ref{mag_T} shows that    the maximum in $\Delta\chi^\prime_{mmt}$ at $T^*$ drops quickly with increasing 
    pressure,  even at pressures well below $P_c$. 
This would mean that as the quantum critical end point is approached, 
   the metamagnetic quantum criticality is not dominated by long wavelength magnetic 
   fluctuations as would be 
   naively expected if the uniform magnetization density is the order 
   parameter for the metamagnetic transition.  
In other words, the metamagnetic 
    transition near the QCEP does not seem to correspond to field-induced 
    ferromagnetism, rather the important fluctuations near the QCEP may 
    be at short wavelength, or they may not be magnetic at all.  
A possible scenario is that the first-order jump 
    in the magnetization near the QCEP could arise from the sudden 
    disappearance of antiferromagnetic correlations, rather than entry 
    into a field-induced ferromagnetic state. 
This may be consistent with the suggestion that the nematic phase is a
    spatially modulated magnetic state as predicted 
    in Ref. \cite{Berridge09, Berridge10}.

In high-purity crystals, field-angle tuned measurements also observed that 
   $\Delta\chi^\prime_{mmt}(T^*)$ drops dramatically as the QCEP is approached \cite{Grigera03}. 
It was suggested that the expected divergence of $\chi$  at $T^*$ was being 
    suppressed by impurity-enhanced critical slowing down, so that 
    the finite frequency ($\sim$80~Hz) used in these ac susceptibility measurements is
    not a good approximation to the zero-frequency limit, and therefore the genuine
    divergence in the long-wavelength limit was not unveiled \cite{Grigera03}. 
However, because we used ultra-pure crystals, with five-times lower residual resistivity, and a 
    significantly lower measurement frequency ($\sim$14 Hz), 
    we feel that it is unlikely that the susceptibility would diverge, even if it were 
    measured at zero frequency.  
This is further supported by our observation that the frequency dependence of the 
    relative variation of $\Delta\chi'_{mmt}$ is extremely weak: 
    at 13.4~kbar, $\Delta\chi^\prime_{mmt}$ vs $T$ shows almost no difference between 
    83~Hz and 14~Hz (see Figure \ref{mag_T}(b)).

Note that pressure inhomogeneity also cannot account 
   for the suppression of the peak in $\chi$ at $T^*$.  
In our measurements we have some indication of pressure inhomogeneity from 
   the width of the superconducting transition of the tin wire used as a 
   pressure gauge, and from the width of the peaks in $\chi$. 
From these we know that the pressure inhomogeneity is very small, as expected for 
   the pressure medium, Daphne oil 7373, at this pressure \cite{Yokogawa07}. 
Moreover, at a given pressure, inhomogeneity in the pressure would 
   broaden the peaks in $\chi$ at all temperatures, so we would still 
   expect to see some enhancement of $\Delta \chi'_{mmt}$ at $T^*$, if such 
   a maximum in $\Delta \chi^\prime_{mmt}$ were present with homogeneous pressure, 
   even if the divergence is partially suppressed; 
   what we actually observe is that the maximum disappears almost completely as the QCEP 
   is approached.  

\begin{figure}
 \includegraphics[width=0.45\textwidth]{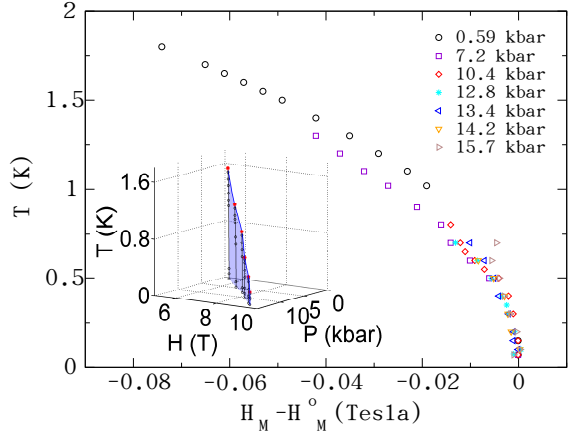}
 \caption{Main figure: the temperature dependence of the metamagnetic transition
    field $H_M$ for each pressure as viewed in the ($H$,$T$) plane, 
    relative to the 70 mK value $H_M^\circ$ at each pressure. 
  Inset: $H_M$ 
    viewed in the ($P,H,T$) space without offset
    to show that the curvature of the metamagnetic `wing' is rather small, 
    although on a fine scale, as shown in the main figure, 
    it is clearly visible.
}
\label{HmvsT}
\end{figure}

The temperature dependence of $H_M$ at different fixed pressures, as shown in 
  Figure \ref{HmvsT}, could also be interpreted as evidence of the importance of 
  quantum fluctuations at finite $q$, or higher-order correlations in the electron system. 
The decrease of $H_M$ with increasing temperature, which is at
  first sight surprising within a simple picture of metamagnetism, 
  has in the past been explained as arising from a growth of quantum 
  fluctuations at long wavelength with decreasing temperature, 
  although Berridge has recently shown that similar curves are 
  generated within a Stoner theory \cite{Millis02, Berridgearxiv10}.  
In  either  scenario, 
    however, one might expect the curvature of $H_M$ to change 
    at $P_c$, whereas we find that the curvature of $H_M$ at $P_c$ is the 
    same as at higher and lower pressures far from $P_c$. 

Finally, our argument that the quantum critical
 fluctuations at the pressure-tuned
 QCEP are not ferromagnetic in nature 
 is supported by other measurements at 
 the field-angle tuned QCEP. Ambient pressure neutron and
NMR studies \cite{Ramos08, Capogna03,Kitagawa05} show that antiferromagnetic fluctuations prevail over ferromagnetic
at low temperatures ($<20$~K). In particular, inelastic neutron measurements
reported by Ramos et al.\ show that, for $H \parallel c$, antiferromagnetic
fluctuations are present in a wide field range (4--13~T), and become soft at
the metamagnetic field \cite{Ramos08}. The NMR study reported by Kitagawa et al.\ further
points out that the quantum critical fluctuations at the quantum critical
point of Sr$_3$Ru$_2$O$_7$ are antiferromagnetic \cite{Kitagawa05}. The finite-$q$ magnetic fluctuations
may be associated with the spatially modulated magnetic phase, i.e the
LOFF nematic phase, which is suggested to exist near the QCEP by Berridge
et al.\ \cite{Berridge09, Berridge10}. The short-range correlations of the LOFF phase may be present
outside of this phase and gain strength as the QCEP is approached \cite{Raghu09}; this
scenario may explain the disappearance of the sharpness of the peak in $\Delta\chi^\prime_{mmt}$
vs $T$ with increasing pressure.

Although pressure tuning for $H \parallel ab$ causes $T^*$ to go 
    smoothly to 0~K, we do see different behavior emerging near the QCEP.
Firstly, there is a change of the slope $dT^*/dP$ at $\sim$12.8~kbar 
  (Figure \ref{phasedia}), indicating a change in the underlying physics. 
Secondly, there is the secondary maximum that appears on the right of the 
    main peak in $\Delta \chi^\prime$ (see Figure \ref{12.8_13.4_14.2kbar}).  
This is present only in 
    the region 12.8 to 16.7~kbar, that is, only near $P_c$, and is 
    reminiscent of the double transition that encloses the nematic phase in 
    the field-tuning measurements. We do not, however, observe a corresponding 
    peak in $\Delta \chi^{\prime\prime}$ at this secondary maximum.  Thirdly, there is a 
    secondary maximum in $\Delta \chi^\prime$ just below the main peak, that may 
    correspond to a weakly split structure in $\Delta \chi^{\prime\prime}$  
    which starts from $\sim$7~kbar and  becomes clear at 13.4~kbar 
    (see Figure \ref{12.8_13.4_14.2kbar}). 
This is a very weak splitting, which we could only resolve by averaging many repeated 
    runs, and the field interval is much smaller
    than is seen for the field-tuned nematic phase: $\sim$0.027~T as opposed to $\sim$0.25~T.

It should be noted that it may be possible to have the nematic phase without
the bounding first-order transitions: the top of the nematic `dome' is defined by
a second-order transition (Figure \ref{genericpha}(ii)). Perhaps, under some conditions, only
the top of the dome exists.  In fact, because the field is being applied in the $ab$-plane so that 
     the in-plane symmetry is already broken, there may be no need for even 
     a second-order phase transition, and it may be possible to enter the 
     nematic state via a crossover.

At this stage, evidence for the nematic phase is not conclusive, and it will
be important to carry out magnetotransport studies near $P_c$, as peaks in
 $\rho (B)$ at low temperature provide definitive evidence for the nematic phase  \cite{Grigera04}.
The only previous hydrostatic pressure study of the 
  magnetoresistance of Sr$_3$Ru$_2$O$_7$ with $H\parallel ab$ was carried out on a 
  high-purity sample at $T=2.5$~K in the pressure range 0 to $\sim 10$~kbar \cite{Chiao02}. 
This study showed a broad magnetoresistance peak around the metamagnetic transition 
  moving  to higher field with increasing pressure at a rate consistent with our observations; 
  however, because the magnetoresistance was measured at a temperature well above $T^*$, 
  and pressures well below $P_c=13.6$, and on  a sample which is not believed to be pure 
  enough to exhibit the nematic phase, no conclusion  can be drawn about the existence of 
  the nematic phase from this work. 

Finally, we address the issue of magnetovolume effects, which are known to play
an important role in  metamagnetism \cite{Chiaounpublish}. For instance, in CeRu$_2$Si$_2$ magnetovolume
effects provide positive feedback to drastically sharpen what would
be a broad crossover under constant volume \cite{Flouquet95,Chiaounpublish}. 
In our measurements, the freezing
of the pressure medium (Daphne oil 7373) at low temperatures ($\sim$200~K)
may suppress positive magnetoelastic feedback in Sr$_3$Ru$_2$O$_7$ -- a system with a
strong magnetoelastic coupling (the magnetic Gr$\rm \ddot{u}$neisen parameter $\Gamma_H >$ 100)\cite{Gegenwart06}. This may broaden the peak in $\Delta\chi^\prime$, connecting the secondary maxima and the central peak to produce weak 
    `shoulders' rather than distinct separate peaks.   We  point
out that some features observed around $P_c$ disappear at higher pressures, for
  instance the secondary maximum to the right of the main peak in 
    $\Delta \chi^\prime$, so they are unlikely to be caused by pressure inhomogeneity in the transmitting medium.

\section{Summary}

In Sr$_3$Ru$_2$O$_7$, it has been previously established that a QCEP can be produced
by tuning the magnetic-field-angle from the $ab$-plane toward the $c$-axis at ambient pressure,
and that in an ultra-pure sample this QCEP is avoided by the appearance of a nematic phase bounded  by two first-order MMTs. In this work, we have used  ac susceptibility
measurements to show that,  for $H \parallel ab$,  hydrostatic pressure  can also
produce a QCEP in an ultra-pure sample. We see that the critical end-point
temperature of the first-order metamagnetic transition $T^*$ falls monotonically
as a function of pressure, going to zero rather suddenly above 12.8~kbar; the
QCEP exists at $P_c = 13.6 \pm 0.2$~kbar. The signature of the nematic phase
observed in field-angle tuning -- two clearly resolved MMTs at the phase 
boundaries -- is absent. 
We also observe that with increasing pressure the divergence of the
   susceptibility at the critical point diminishes quickly, 
   suggesting that short-wavelength fluctuations 
   may dominate the metamagnetic transition as the QCEP is approached.



\end{document}